\documentclass[12pt,thmsa]{article}
\usepackage{amssymb}

\input tcilatex
\begin{document}

\begin{center}
\textbf{EXCITATION OF NONLINEAR TWO-DIMENSIONAL WAKE WAVES IN
RADIALLY-NONUNIFORM PLASMA}

\smallskip\ 

Arsen G. Khachatryan

\smallskip 

\textit{Yerevan Physics Institute, Alikhanian Brothers Street 2, }

\textit{Yerevan 375036, Armenia}

\smallskip 
\end{center}

It is shown that an undesirable curvature of the wave front of
two-dimensional nonlinear wake wave excited in uniform plasma by a
relativistic charged bunch or laser pulse may be compensated by radial
change of the equilibrium plasma density.

\smallskip 

PACS number(s): 52.40.Mj, 52.40.Nk, 52.75.Di, 52.35.Mw

\vspace{1.0in}

The progress in the technology of ultrahigh intensity lasers and high
current relativistic charged bunch sources permits the use of laser pulses
[1] or charged bunches [2] for excitation of strong plasma waves. The
excited plasma waves can be used both for acceleration of charged particles
and focusing of bunches to get high luminosity in the linear colliders [2].
At present the plasma-based accelerator concepts are actively developed both
theoretically and experimentally (see the overview in Ref. [3] and
references therein).

The amplitude of longitudinal electric field $E_{\max }$ in relativistic
wake-waves excited in cold plasma is limited by the relativistic
wave-breaking field [4] $E_{rel}=[2(\gamma -1)]^{1/2}E_{WB}/\beta $, where $%
\gamma =(1-\beta ^2)^{-1/2}$ is a relativistic factor, $\beta =v_{ph}/c$ is
a dimensionless phase velocity of the wave, $E_{WB}=m_e\omega _{pe}v_{ph}/e$
($E_{WB}[V/cm]\approx 0.96n_p^{1/2}[cm^{-3}]$) is the conventional
nonrelativistic wave-breaking field, $\omega _{pe}=(4\pi n_pe^2/m_e)^{1/2}$
is the electron plasma frequency,\textbf{\ } $n_p$ is the equilibrium
density of plasma electrons, $m_e$ and $e$ are the mass and absolute value
of the electron charge. The acceleration rate in the wake fields can reach
tens of $GeV/m$, that much exceeds the rates reached in conventional
accelerators.

The linear wake-field theory is valid when $E_{\max }\ll E_{WB}$. In case of
wake wave excitation by relativistic charged particle bunch [Plasma
Wakefield Accelerator (PWFA)] this corresponds to the condition $\alpha
=\left| q\right| n_b\diagup en_p\ll 1$ [2], where $q$ is the charge of bunch
particles, $n_b$ is their concentration. In the scheme of wake wave
excitation by a short laser pulse (of the length comparable with plasma
wavelength $\lambda _p=2\pi v_{ph}\diagup \omega _{pe}$; Laser Wakefield
Accelerator (LWFA)), the linear theory is valid when $a^2=e^2E_0^2\diagup
(m_e^2c^2\omega _0^2)\ll 1$ [5], where $\omega _0\gg \omega _{pe}$ and $E_0$
are the frequency and amplitude of laser radiation respectively. The phase
velocity of the wake wave is equal to the bunch velocity $v_b$ in PWFA and
to the laser pulse group velocity $v_g\approx c(1-\omega _{pe}^2/2\omega
_0^2)$ ( that corresponds to $\gamma =\gamma _g\approx \omega _0/\omega
_{pe}\gg 1$) in LWFA.

The one-dimensional nonlinear wake waves excited by wide drivers (when $%
k_pr_d\gg 1$, where $k_p=\omega _{pe}/v_{ph}$ is the wavenumber and $r_d$ is
the radius of the driver) are studied in sufficient detail both for PWFA\
[6] and for LWFA\ [7]. These studies testify to the feasibility of
excitation of strong nonlinear plasma waves with the amplitude of up to $%
E_{rel}$ by bunches with $\alpha \gtrsim 0.5$ and laser pulses with $%
a^2\gtrsim 1$. The other important results of the one-dimensional nonlinear
theory are the steepening of the wake wave and the increase of wavelength
with amplitude. The wave with the amplitude $E_{\max }\approx E_{rel}$ has
the wavelength nearly $\gamma ^{1/2}$ times as large as the linear
wavelength $\lambda _p$.

In reality, the transverse sizes of the drivers used are ordinarily
comparable or less than their longitudinal size. The allowance for finite
transverse sizes of the drivers and, accordingly, the transverse motion of
plasma electrons complicate the treatment of the problem in the nonlinear
regime. In the general case the analytical solution of this regime seems
impossible and here the use of numerical methods are usually required. The
numerical investigation of nonlinear effects in two-dimensional
(axially-symmetrical) wake waves exited in uniform plasma has shown that in
the nonlinear regime the wavelength becomes dependent on the transverse
(relative to the driver propagation direction) coordinate [3,8-10]. The
change of a nonlinear wavelength in the transverse direction is due to the
dependence of wavelength on the amplitude, that, in its turn, is varied in
the radial direction owing to finite cross section of the driver. This leads
to a curvature of the phase front of the nonlinear wave [8-10], to
steepening and ''oscillations'' of the field in the transverse direction
[10,11] and eventually to the development of turbulence. From the viewpoint
of acceleration and focusing of charged bunches in the wave, the curvature
of the nonlinear wave front is undesired as the quality (emittance,
monochromaticity) of the driven bunch worsens. In the present work we show
that by means of wake wave excitation in plasma, the density of which is
properly varied in the transverse direction, one can eliminate the nonlinear
change of wavelength in the transverse direction and the related curvature
of phase front. The plasma of this kind may be produced by charged beams
[12] or laser pulses [13] passing through a neutral gas or a partially
ionized uniform plasma due to an additional ionization.

Equations for non-zero components of plasma electrons momentum and
electromagnetic field describing the steady nonlinear wake-fields in
radially-nonuniform plasma can be obtained by simple generalization of
equations for uniform plasma [10,11,14,15]:

\begin{equation}
\beta \frac{\partial P_z}{\partial z}-\frac{\partial \gamma _e}{\partial z}%
-\beta ^2E_z=0,  \tag{1}
\end{equation}

\begin{equation}
\beta \frac{\partial P_r}{\partial z}-\frac{\partial \gamma _e}{\partial r}%
-\beta ^2E_r=0,  \tag{2}
\end{equation}
\begin{equation}
-\frac{\partial H_\theta }{\partial z}+\beta \frac{\partial E_r}{\partial z}%
+\beta _rN_e=0,  \tag{3}
\end{equation}
\begin{equation}
\nabla _{\bot }H_\theta +\beta \frac{\partial E_z}{\partial z}+\beta
_zN_e+\beta \alpha =0,  \tag{4}
\end{equation}
\begin{equation}
\beta \frac{\partial H_\theta }{\partial z}-\frac{\partial E_r}{\partial z}+%
\frac{\partial E_z}{\partial r}=0,  \tag{5}
\end{equation}
\begin{equation}
N_e=N_p(r)-\alpha -\nabla _{\bot }E_r-\frac{\partial E_z}{\partial z}. 
\tag{6}
\end{equation}
As usual, Eqs. (1) and (2) were derived taking into account the conservation
of generalized momentum $\beta ^2\mathbf{H}-rot\mathbf{P}=0$, or in our case 
\begin{equation}
\beta ^2H_\theta +\frac{\partial P_z}{\partial r}-\frac{\partial P_r}{%
\partial z}=0.  \tag{7}
\end{equation}
In Eqs. (1) - (6) $\gamma _e=(1+P_z^2+P_r^2+a^2/2)^{1/2}$ , $\beta _{z,\text{
}r}=P_{z,\text{ }r}$ $/\gamma _e$ and $N_e=n_e/n_p(r=0)$ are respectively a
relativistic factor, dimensionless components of velocity and dimensionless
density of plasma electrons, $N_p=n_p(r)/n_p(0)$, $\beta =v_{ph}/c$, $%
z=k_p(r=0)(Z-v_{ph}t)_{\text{, }}\nabla _{\bot }=\partial /\partial r+1/r.$
Also the following dimensionless variables have been used: the space
variables are normalized on $\lambda _p(r=0)/2\pi =1/k_p(r=0)$, the momenta
and velocities - respectively on $m_ec$ and the velocity of light and the
strengths of electric and magnetic fields - on the nonrelativistic
wave-breaking field at the axis $E_{WB}(r=0)=m_e\omega _{pe}(r=0)v_{ph}/e$.
The field of forces acting on relativistic electrons in the excited field is 
$\mathbf{F}(-eE_z,-e(E_r-\beta H_\theta ),0)$. In PWFA $\alpha \neq 0$, $%
a^2=0$ and in LWFA $\alpha =0$, $a^2\neq 0$ (in Eqs. (1)-(6) the linear
polarization of the laser pulse field is assumed; for the circular
polarization the value of $a$ should be multiplied by the factor $2^{1/2}$).

We have solved Eqs. (1)-(6) numerically choosing the Gaussian profile of the
driver both in longitudinal and transverse directions:

\begin{equation}
A(z,r)=A_0\exp [-(z-z_0)^2/\sigma _z^2]\exp (-r^2/\sigma _r^2),  \tag{8}
\end{equation}
where $A(z,r)$ stands for $\alpha =n_b(z,r)/n_p(r=0)$ or $a^2(z,r)$. Shown
in Fig. 1 is the nonlinear 2D plasma wake wave excited in uniform plasma [$%
N_p(r)=1$] by the relativistic electron bunch ($\alpha _0=0.4$, $\sigma _z=2$%
, $\sigma _r=5$; for example, in this case $n_{b0}=4\times 10^{13}cm^{-3}$
and the characteristic longitudinal and transverse sizes of the bunch $%
\sigma _{z,r}/k_p$ correspondingly are $1.06mm$ and $2.65mm$ when $%
n_p=10^{14}cm^{-3}$). One can see that the wavelength changes with the
radial coordinate $r$. This leads to curving of the phase front and to
''oscillations'' in the transverse direction (see Fig. 2, curve 1). As $%
\left| z\right| \,\,\,$increases, the change of phase in transverse
direction (for fixed $z$) becomes more and more marked. The longitudinal
space parameter characterizing the nonlinear wave front curving is [10]:

\begin{equation}
\xi =\frac{\lambda _p}{2[1-\lambda _p/\Lambda (0)]},  \tag{9}
\end{equation}
where $\Lambda (r)$ is the nonlinear wavelength. At the distance $\left|
\Delta z\right| \thickapprox \xi $ from the driver the oscillation phase at
the axis ($r=0$) is opposite to that on the periphery ($r\gtrsim \sigma _r$%
). Thus, in 2D nonlinear regime the nonlinear wavelength changes with $r$
due to nonlinear increase of the wavelength with wave amplitude. On the
other hand, the linear wavelength $\lambda _p\sim n_p^{-1/2}$ decreases with
equilibrium density of plasma. Hence follows an opportunity to compensate
the nonlinear increase in wavelength by reducing the wavelength that is due
to the growth of equilibrium density of plasma. Indeed, assume that the
nonlinear wavelength of the two-dimensional wake wave in the uniform plasma $%
\Lambda (r)$ is known. Then, one can roughly compensate for the radial
variation of the nonlinear wavelength by changing the equilibrium density of
plasma in the radial direction according to the relation

\begin{equation}
\Lambda (0)/\Lambda (r)=\lambda _p(r)/\lambda _p(0)=[n_p(0)/n_p(r)]^{1/2}. 
\tag{10}
\end{equation}
In this case the equation for equiphase surfaces is $z\approx const$, and,
therefore, the solution for fields could be written in the form $%
f_1(z)f_2(r) $. If we put that the function $\Lambda (r)$ is Gaussian (that
is approximately the case at least for $r<\sigma _r,$ according to numerical
data for profiles (8)), then one can take the transverse profile of the
equilibrium plasma density to be also Gaussian:

\begin{equation}
n_p(r)=n_{p0}\exp (-r^2/\sigma _p^2).  \tag{11}
\end{equation}
Then follows from Eqs. (10) and (11) that

\begin{equation}
\sigma _p=r/[\ln (\Lambda (0)/\Lambda (r))]^{1/2}.  \tag{12}
\end{equation}
For example, according to (12), the numerical data for $\Lambda (r)$ in the
nonlinear wave shown in Fig.1 give $\sigma _p\approx 11$. Thus, in the
radially nonuniform plasma, the density of which is changed according to
(11) and (12), one can practically avoid undesirable curvature of the wave
front of a nonlinear wave. Figs. 3 and 4 illustrate the validity of this
assertion respectively for PWFA and LWFA (see also Fig.2, curve 2). One can
see, that the nonlinear wavelength in the nonuniform plasma with proper
radial profile does not practically change in the transverse direction.

As for the case of LWFA, one has to note that as is well known, without
optical guiding the diffraction limits the distance of laser-plasma
interaction (and, hence, the energy gain of particles accelerated by a wake
wave) to a few Reyleigh lengths $Z_R=\pi r_0^2/\lambda $, where $r_0$ the
minimum laser pulse spot size at the focal point and $\lambda $ is the laser
wavelength. For high-intensity laser pulses the quantity $Z_R$ is usually of
the order of several millimeters. One of approaches in preventing of
diffraction broadening of laser pulse and increasing of laser-plasma
interaction distance is the guiding of the pulse in preformed plasma density
channel [3]. Here the unperturbed plasma density grows from the pulse axis ($%
r=0$) to its periphery, in contrast with the radial profile of the plasma
density that was proposed above for prevention of phase front curvature of a
nonlinear wake wave when the plasma density at the driver axis is maximum.
In case of nonlinear wake wave excitation by high power laser pulses (of
power $P>P_c=2c(e/r_e)^2[\omega _0/\omega _{pe}(r=0)]^2\approx 17[\omega
_0/\omega _{pe}(r=0)]^2$ $GW$, where $r_e=e^2/m_ec^2$ is the classical
electron radius), the process of diffraction broadening of a larger part of
pulse, including the case of proposed equilibrium profile of plasma, may be
prevented or essentially retarded due to the relativistic self-focusing [3].
One can estimate the condition of relativistic self-focusing in our case
from the following relation (see, e. g. Sec. VI in Ref. [3]): $%
P/P_c>1+(\Delta n/\Delta n_c)(r_s/r_0)^4$, where $r_s$ is the spot size, $%
\Delta n_c=1/\pi r_er_0^2$, $\Delta n\simeq n_p(0)-n_p(\sigma _r)$ is the
radial variation of unperturbed plasma density in the laser pulse.

This work has been supported by the International Science and Technology
Center under Project No. A-013.

\begin{center}
\textbf{REFERENCES}
\end{center}

[1] \thinspace T. Tajima and J. M. Dawson, Phys. Rev. Lett. \textbf{43}, 267
(1979).

[2] \thinspace R. D. Ruth, A. W. Chao, P. L. Morton, and P. B. Wilson, Part.
Accel. \textbf{17}, 171 (1985); P. Chen, Part. Accel. \textbf{20}, 171
(1987).

[3] $\,$E. Esarey, P. Sprangle, J. Krall, and A. Ting, IEEE Trans. Plasma
Sci. \textbf{24}, 252 (1996).

[4] \thinspace A. I. Akhiezer and R .V. Polovin, Zh. Eksp. Teor. Fiz. 
\textbf{30}, 915 (1956) [Sov. Phys. JETP \textbf{3}, 696 (1956)].

[5] $\,$L. M. Gorbunov and V. I. Kirsanov, Sov. JETP \textbf{66}, 290 (1987).

[6] \thinspace A. Ts. Amatuni, E. V. Sekhpossian, and S. S. Elbakian, Fiz.
Plasmy \textbf{12}, 1145 (1986); \thinspace J. B. Rosenzweig, Phys. Rev.
Lett.\textbf{\ 58}, 555 (1987); \thinspace A. Ts. Amatuni, S. S. Elbakian,
A. G. Khachatryan, and E. V. Sekhpossian, Journal of Contemporary Physics
(Allerton Press, Inc., NY) \textbf{28}, 8 (1993); A. G. Khachatryan, Phys.
Plasmas\textbf{\ 4}, 4136 (1997).

[7] S. V. Bulanov, V. I. Kirsanov, and A. S. Sakharov, JETP Lett. \textbf{50}%
, 198 (1989); P. Sprangle, E. Esarey, and A. Ting, Phys. Rev. A \textbf{41},
4463 (1990); V. I. Berezhiani and I. G. Murusidze, Phys. Lett. A \textbf{148}%
, 338 (1990).

[8] C. D. Decker, W. B. Mori, and T. Katsouleas, Phys. Rev. E \textbf{50},
3338 (1994).

[9] S. V. Bulanov, F. Pegoraro, and A. M. Pukhov, Phys. Rev. Lett. \textbf{74%
}, 710 (1995).

[10] A. G. Khachatryan and S. S. Elbakian, Proceedings PAC'99, New York,
1999.

[11] B. N. Breizman, T. Tajima, D. L. Fisher, and P. Z. Chebotaev, In: 
\textit{Research Trends in Physics: Coherent Radiation and Particle
Acceleration , edited by A. Prokhorov }(American Institute of Physics, New
York, 1992), pp. 263-287.

[12] A. K. Berezin et al., Plasma Phys. Rep. \textbf{20}, 596 (1994).

[13] V. Malka et al., Phys. Rev. Lett. \textbf{79}, 2979 (1997).

[14] P. Mora and T. M. Antonsen Jr., Phys. Plasmas \textbf{4}, 217 (1997).

[15] K. V. Lotov, Phys. Plasmas \textbf{5}, 785 (1998).

\newpage\ 

\begin{center}
\textbf{FIGURE CAPTIONS}
\end{center}

Fig. 1. The two-dimensional nonlinear wake wave in uniform plasma [$N_p(r)=1$%
]. The parameters of the bunch are:$\,\alpha _0=0.4$, $\sigma _z=2$,$%
\,\,\sigma _r=5$, $\gamma =10$. (a). The density of plasma electrons $N_e$
and of the bunch. 1 --- the density of plasma electrons at the axis, $r=0$;
2 --- the same for $r=2$; 3 --- $r=4$; 4 --- $r=5$; 5 --- the density of
bunch at the axis $\alpha (z,r=0)$. (b). The longitudinal electric field for 
$r=0,2,4$ and $5$ in the order of magnitude reduction. (c). The focusing
field $f_r=\beta H_\theta -E_r$. 1 --- $r=2$; 2 --- $r=4$; 3 --- $r=5$. All
variables are normalized.

Fig. 2. The radial behavior of the normalized longitudinal electric field
strength $E_z$. 1 --- $E_z(z=-25,r)$ in the nonlinear wake wave excited in
uniform plasma for the case given in Fig. 1 ($\left| \Delta z\right|
\thickapprox \xi $ [see Eq. (9)]); 2 --- $E_z(z=-25,r)$ in nonuniform plasma
for the case given in Fig. 3.

Fig. 3. The two-dimensional nonlinear wake wave in nonuniform plasma with $%
\sigma _p=11$. The bunch parameters are the same as in Fig. 1. 1 --- the
density of plasma electrons versus $z$ for $r=0,2,4$ and $5$ in the order of
magnitude reduction. (b). The same for the longitudinal electric field. (c).
The focusing field. 1 --- $r=2$; 2 --- $r=4$; 3 --- $r=5$. All variables are
normalized.

Fig. 4. The two-dimensional nonlinear wake wave excited by laser pulse. The
pulse parameters are : $a_0^2=3.6$, $\sigma _z=2$,$\,\,\sigma _r=5$, $\gamma
=10$. (a). The dimensionless accelerating field $E_z$ excited in uniform
plasma. $r=0,2,4$ and $5$ in the order of magnitude reduction. (b). The same
in the nonuniform plasma, $\sigma _p=12$.

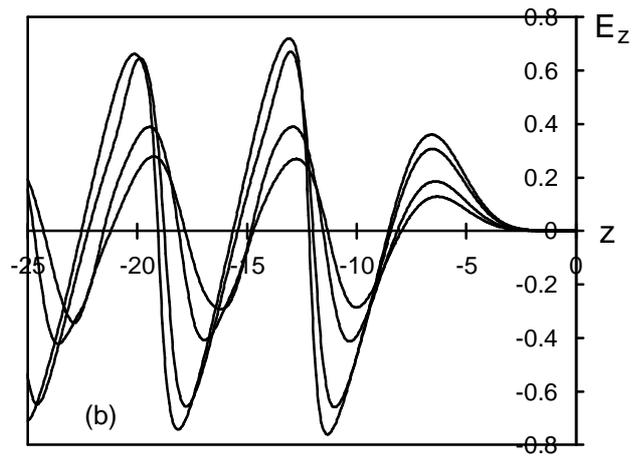
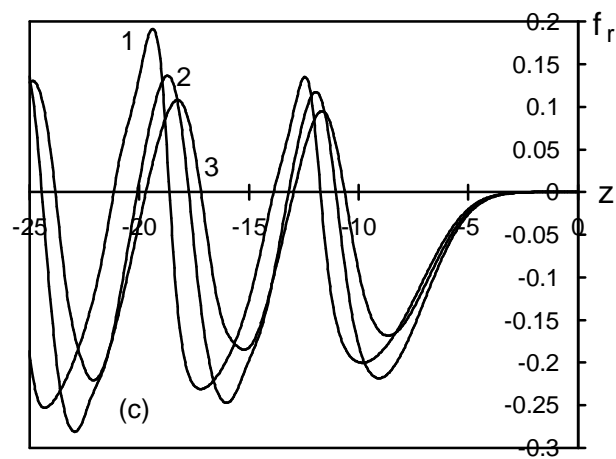
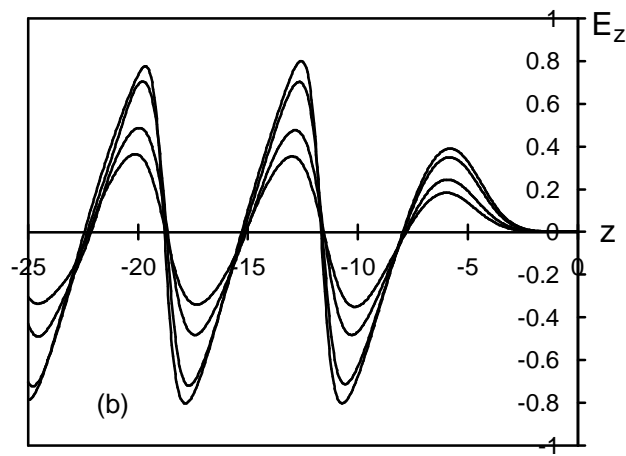
\end{document}